\documentclass{bioinfo}
\copyrightyear{2003}
\pubyear{2003}

\begin{document}
\firstpage{1}

\title[Graphlet Degree Distribution]{Biological Network Comparison Using Graphlet Degree Distribution}
\author[Pr\v{z}ulj, N.]{Nata\v{s}a Pr\v{z}ulj}
\address{Computer Science Department, University of California, Irvine, CA 92697-3425, USA}

\history{Received on XXXXX; revised on XXXXX; accepted on XXXXX}

\editor{Associate Editor: XXXXXXX}

\maketitle

\begin{abstract}

\section{Motivation:}
Analogous to biological sequence comparison, comparing cellular networks
is an important problem that could provide insight into biological understanding 
and therapeutics.  For technical reasons, comparing large networks is computationally
infeasible, and thus heuristics, such as the degree distribution, clustering coefficient,
diameter, and relative graphlet frequency distribution have been sought.  It is
easy to demonstrate that two networks are different by simply showing a short list
of properties in which they differ.  It is much harder to show that two networks 
are similar, as it requires demonstrating their similarity in \emph{all} of their
exponentially many properties.  Clearly, it is computationally prohibitive to analyze 
all network properties, but the larger the number of constraints we impose in determining
network similarity, the more likely it is that the networks will truly be similar.

\section{Results:} We introduce a new systematic measure of a network's
local structure that imposes a large number of similarity constraints on networks 
being compared.  In particular, we generalize the degree distribution,
which measures the number of nodes ``touching'' $k$ edges, into distributions
measuring the number of nodes ``touching'' $k$ \emph{graphlets}, where graphlets
are small connected non-isomorphic subgraphs of a large network.
Our new measure of network local structure
consists of $73$ \emph{graphlet degree distributions} of graphlets with 2, 3, 4, and 5 nodes,  
but it is easily extendible to a greater number of constraints (i.e, graphlets), if necessary, 
and the extensions are limited only by the available CPU.  Furthermore, we show a way to 
combine the $73$ graphlet degree distributions into
a network ``agreement'' measure which is a number between $0$ and $1$, where $1$ means 
that networks have identical distributions and $0$ means that they are far apart.
Based on this new network agreement measure, we show that almost all of the fourteen 
eukaryotic PPI networks, including human, resulting from various high-throughput 
experimental techniques, as well as from curated databases, are better modeled 
by geometric random graphs than by Erdos-Renyi, random scale-free, or Barabasi-Albert 
scale-free networks.

\section{Availability:} Software executables are available upon request.

\section{Contact:} \href{natasha@ics.uci.edu}{natasha@ics.uci.edu}
\end{abstract}

\section{Introduction}\label{sect:intro}

Understanding cellular networks is a major problem in current
computational biology.  These networks are commonly modeled by \emph{graphs}
(also called \emph{networks}) with \emph{nodes} representing biomolecules such as 
genes, proteins, metabolites etc., and \emph{edges} representing physical, 
chemical, or functional interactions between the biomolecules.
The ability to compare such networks would be very useful.
For example, comparing a diseased cellular network to a healthy one
may aid in finding a cure for the disease, and
comparing cellular networks of different species could enable 
evolutionary insights.  A full description
of the differences between two large networks is infeasible
because it requires solving the {\it subgraph isomorphism} problem,
which is an NP-complete problem.
Therefore, analogous to the BLAST heuristic \citep{Altschul90}  for biological 
sequence comparison, we need to design a heuristic tool for the full-scale 
comparison of large cellular networks \citep{Berg04}.
The current network comparisons consist of heuristics falling into two
major classes:  1) global heuristics, such as counting the number of connections 
between various parts of the network (the ``degree distribution''), computing the average
density of node neighborhoods (the ``clustering coefficients''), or the average
length of shortest paths between all pairs of nodes (the ``diameter''); and 2)
local heuristics that measure relative distance between concentrations of
small subgraphs (called \emph{graphlets}) in two networks \citep{Przulj04}.

Since cellular networks are incompletely explored, global statistics
on such incomplete data may be substantially biased, or even misleading with 
respect to the (currently unknown) full network.  Conversely, certain neighborhoods
of these networks are well-studied, and so locally based statistics applied to
the well-studied areas are more appropriate.
A good analogy would be to imagine that MapQuest knew details of the
streets of New York City and Los Angeles, but had little knowledge of
highways spanning the country.  Then, it could provide good
driving directions inside New York or L. A., but not between the two.
Similarly, we have detailed knowledge of certain local areas of
biological networks, but data outside these well-studied areas is currently
incomplete, and so global statistics are likely to provide misleading
information about the biological network as a whole, while local statistics
are likely to be more valid and meaningful.

Due to the noise and incompleteness of cellular network data, 
local approaches to analyzing and comparing cellular network structure that involve 
searches for small subgraphs have been successful in analyzing, modeling, and 
discovering functional modules in cellular 
networks \citep{Milo02,Shen-Orr02,Milo04,Przulj04}.
Note that it is easy to show that two networks are different simply by finding
any property in which they differ.  However, it is much harder to show that
they are similar, since it involves showing that two networks are similar with
respect to \emph{all} of their properties. 
Current common approaches to show network similarity are based on
listing several common properties, such as the degree distribution, clustering,
diameter, or relative graphlet frequency distribution.  The larger the number
of common properties, the more likely it is that the two networks are similar.
But any short list of properties can easily be mimicked by two very large and
different networks.
For example, it is easy to construct networks with exactly the same degree distribution
whose structure and function differ substantially \citep{Przulj04,Li05,Doyle05}.

In this paper, we design a new local heuristic for measuring network structure
that is a \emph{direct} generalization of the degree distribution.  In fact, the
degree distribution is the first in the spectrum of $73$ 
\emph{graphlet degree distributions} that are components of this new measure of 
network structure.
Thus, in our new network similarity measure, we impose $73$ highly structured 
constraints in which
networks must show similarity to be considered similar; this is a much larger
number of constraints than provided by any of the previous approaches and 
therefore it increases the chances that two networks are truly similar if they
are similar with respect to this new measure.  Moreover, the measure
can be easily extended to a greater number of constraints simply by adding
more graphlets.  The extensions are limited only by the available CPU.

Based on this new measure of structural
similarity between two networks, we show that the geometric random graph model 
shows exceptionally high agreement with twelve out of fourteen different 
eukaryotic protein-protein 
interaction (PPI) networks.  Furthermore, we show that such high structural agreements 
between PPI and geometric random graphs are unlikely to be beaten by another random 
graph model, at least under this measure.

\subsection{Background}\label{sect:background}

Large amounts of cellular network data for a number of organisms
have recently become available through high-throughput 
methods \citep{Ito00,uetz00,GiotSci03,Vidal04,Stelzl05,Rual05}.  
Statistical and theoretical properties of these networks have been extensively
studied \citep{Jeong00,Maslov02,Shen-Orr02,Milo02,Vazquez04,Yeger-Lotem04,Przulj04,Tanaka05} due to their important biological implications \citep{Jeong01,LappeHolm04}.  

Comparing large cellular networks is computationally intensive.
Exhaustively computing the differences between networks is 
computationally infeasible, and thus efficient 
heuristic algorithms have been sought \citep{Kashtan04,Przulj06}. 
Although \emph{global properties} of large networks
are easy to compute,
they are inappropriate for use on incomplete networks because
they can at best describe the
structure produced by the biological sampling techniques used to obtain the
partial networks \citep{Vidal05}.
Therefore, bottom-up or \emph{local} heuristic 
approaches for studying network structure have been 
proposed \citep{Milo02,Shen-Orr02,Przulj04}.
Analogous to sequence motifs, \emph{network motifs} have been defined as 
subgraphs that occur in a network at frequencies much higher than expected at 
random \citep{Milo02,Shen-Orr02,Milo04}.  Network motifs have been generalized 
to \emph{topological motifs} as recurrent ``similar'' network sub-patterns  \citep{Berg04}.
However, the approaches based on network motifs ignore infrequent subnetworks
and subnetworks with ``average'' frequencies, and thus are not sufficient for
full-scale network comparison.  Therefore, small connected non-isomorphic 
induced subgraphs of a large network, called \emph{graphlets}, have been introduced 
to design a new measure of local structural similarity between two networks
based on their relative frequency distributions \citep{Przulj04}.  

The earliest attempts to model real-world networks
include \emph{Erdos-Renyi random graphs} (henceforth denoted by ``ER'') 
in which edges between pairs of nodes are distributed uniformly at 
random with the same probability $p$ \citep{ErdosRenyi59,ErdosRenyi60}.  
This model poorly describes several properties
of real-world networks, including the degree distribution and clustering coefficients, 
and therefore it has been refined into \emph{generalized random graphs}
in which the edges are randomly chosen as in Erdos-Renyi random graphs, but 
the degree distribution is constrained to match the degree distribution of the real 
network (henceforth we denote these networks by ``ER-DD'').  Matching other global 
properties of the real-world networks to the model networks, such as clustering 
coefficients, lead to further improvements in modeling real-world networks including 
\emph{small-world} \citep{Watts-Strogatz98,Newman99a,Newman99b} and 
\emph{scale-free} \citep{Simon55,Barabasi99} network models
(henceforth, we denote by ``SF'' scale-free Barabasi-Albert networks).
Many cellular networks have been described as scale-free \citep{Barabasi_Oltvai04}.
However, this issue has been heavily debated 
\citep{Aguiar05,Stumpf05,Vidal05,Tanaka05}.
Recently, based on the local relative graphlet frequency distribution measure, 
a geometric random graph model \citep{Penrose03} has been proposed for high-confidence PPI 
networks \citep{Przulj04}.

\section{Approach}\label{sect:approach}

In section \ref{sect:data_model}, we describe the fourteen PPI networks and the four
network models that we analyzed.  Then we describe how we generalize the degree 
distribution to our spectrum of graphlet degree distributions (section \ref{sect:gdd});
note that the degree distribution is the first distribution in this spectrum, since
it corresponds to the only graphlet with two nodes.
Finally, we construct a new measure of similarity between two networks based on 
graphlet degree distributions (section \ref{sect:gdd_agree}).  We describe the
results of applying this measure to the fourteen PPI networks in section \ref{sect:results}.

\subsection{PPI and Model Networks}\label{sect:data_model}

We analyzed PPI networks of the eukaryotic organisms yeast \emph{S. cerevisiae},
frutifly \emph{D. melanogaster}, nematode worm \emph{C. elegans}, and human.
Several different data sets are available for yeast and human, so we analyzed five
yeast PPI networks obtained from three different high-throughput 
studies \citep{uetz00,Ito00,Mering02} and five human PPI networks obtained 
from the two recent high-throughput studies \citep{Stelzl05,Rual05} and three 
curated data bases \citep{Bader03,hprd,MINT}.  We denote by ``YHC'' the high-confidence
yeast PPI network as described by \citealp{Mering02}, by
``Y11K'' the yeast PPI network defined by the top $11,000$ interactions in the
\citealp{Mering02} classification, by ``YIC'' the \citealp{Ito00} 
``core'' yeast PPI network, by ``YU'' the \citealp{uetz00} yeast PPI
network, and by ``YICU'' the union of \citealp{Ito00} core and
\citealp{uetz00} yeast PPI networks (we unioned them as did \citealp{Vidal05} 
to increase coverage).  ``FE'' and ``FH'' denote the fruitfly
\emph{D. melanogaster} entire and high-confidence PPI networks published 
by \citealp{GiotSci03}.  Similarly, ``WE'' and ``WC'' denote the
worm \emph{C. elegans} entire and ``core'' PPI networks published by 
\citealp{Vidal04}.  Finally, ``HS'', ``HG'', ``HB'', ``HH'', and ``HM''
stand for human PPI networks by \citealp{Stelzl05}, 
\citealp{Rual05}, from BIND \citep{Bader03}, HPRD \citep{hprd}, and MINT \citep{MINT}, 
respectively (BIND, HPRD, and MINT data have been downloaded from OPHID \citep{Brown05} on
February 10, 2006).  Note that these PPI networks come from of a wide array of experimental
techniques; for example, YHC and Y11K are mainly coming from tandem affinity 
purifications (TAP) and high throughput MS/MS protein complex identification 
(HMS-PCI), while YIC, YU, YICU, FE, FH, WE, WH, HS, and HG are yeast two-hybrid (Y2H), 
and HB, HH, and HM are a result of human curation (BIND, HPRD, and MINT).

The four network models that we compared against the above fourteen PPI networks
are ER, ER-DD, SF, and 3-dimensional geometric random graphs (henceforth denoted 
by ``GEO-3D'').  Model networks corresponding to a PPI network have the same number
of nodes and the number of edges within $1\%$ of the PPI network's (details of
the construction of model networks are presented by \citealp{Przulj04}). For each of the
fourteen PPI networks, we constructed and analyzed $25$ networks belonging to each
of these four network models.  Thus, we analyzed the total of 
$14 + 14\cdot4\cdot25 = 1,414$ networks.
We compared the agreement of each of the fourteen PPI networks 
with each of the corresponding $4\cdot25=100$ model networks described above 
(our new agreement measure is described in section \ref{sect:gdd_agree}).  
The results of this analysis are presented in section \ref{sect:results}.

\subsection{Graphlet Degree Distribution (GDD)}\label{sect:gdd}

\begin{figure}[ht]
        \centering
        \scalebox{.35}{\includegraphics{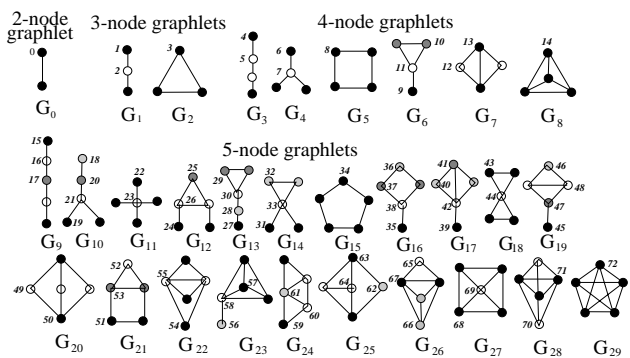}}
        \caption[Graphlet automorphism orbits]{{Automorphism orbits $0,1,2, \ldots, 72$ for the thirty 2, 3, 4, and 5-node graphlets $G_0, G_1, \ldots, G_{29}$.  In a graphlet $G_i$, $i \in \{0,1,\ldots29\}$, nodes belonging to the same orbit are of the same shade.}}
        \label{fig:graphlet_orbits}
\end{figure}

We generalize the notion of the degree distribution as follows.
The degree distribution measures, for each value of $k$,  
the number of nodes of degree $k$.  In other words, for each value of $k$,
it gives the number of nodes ``touching'' $k$ edges.  Note that 
\emph{an  edge} is the only \emph{graphlet with two nodes}; henceforth, 
we call this graphlet $G_0$ (illustrated in Figure \ref{fig:graphlet_orbits}).  
Thus, the degree distribution measures the following: how many nodes ``touch'' one 
$G_0$, how many nodes ``touch'' two $G_0$s, $\ldots$, how many nodes ``touch'' $k$ $G_0$s.
Note that there is nothing special about graphlet $G_0$ and that there is no reason
not to apply the same measurement to other graphlets.  Thus, in addition to applying this
measurement to an edge, i.e., graphlet $G_0$, as in the degree distribution, we apply it to 
the twenty-nine graphlets $G_1, G_2, \ldots G_{29}$ presented in Figure 
\ref{fig:graphlet_orbits} as well.

When we apply this measurement to graphlets $G_0, \ldots, G_{29}$, 
we need to take care of certain topological issues that we first illustrate in
the following example and then define formally.  For graphlet $G_1$, we ask how many
nodes touch a $G_1$; however, note that it is topologically relevant to 
distinguish between nodes touching a $G_1$ at an end or at the middle node.  
This is due to the following mathematical property of $G_1$:
a $G_1$ admits an automorphism (defined  below) that maps its end 
nodes to each other and the middle node to itself.  To understand this phenomenon,
we need to recall the following standard mathematical definitions.  
An \emph{isomorphism} $g$ from graph $X$ to graph $Y$ is a bijection 
of nodes of $X$ to nodes of $Y$ 
such that $xy$ is an edge of $X$ if and only if $g(x)g(y)$ is an edge of $Y$; an 
\emph{automorphism} is an isomorphism from a graph to itself.  The automorphisms
of a graph $X$ form a \emph{group}, called the \emph{automorphism group of} $X$,
and commonly denoted by $Aut(X)$.  If $x$ is a node of graph $X$, then the
\emph{automorphism orbit} of $x$ is 
$Orb(x) = \{y \in V(X) | y=g(x) \text{ for some } g \in Aut(X)\}$, where $V(X)$
is the set of nodes of graph $X$.  
Thus, end nodes of a $G_1$ belong to one automorphism orbit,
while the mid-node of a $G_1$ belongs to another.  Note that graphlet $G_0$ (i.e., an
edge) has only one automorphism orbit, as does graphlet $G_2$; graphlet $G_3$ has two
automorphism orbits, as does graphlet $G_4$, graphlet $G_5$ has one automorphism orbit, 
graphlet $G_6$ has three automorphism orbits etc. (see Figure \ref{fig:graphlet_orbits}).
In Figure \ref{fig:graphlet_orbits}, we illustrate the partition of nodes of 
graphlets $G_0, G_1, \ldots, G_{29}$ into automorphism orbits
(or just \emph{orbits} for brevity); henceforth, we number the $73$ 
different orbits of graphlets $G_0, G_1, \ldots, G_{29}$ from $0$ to $72$,
as illustrated Figure \ref{fig:graphlet_orbits}.  Analogous to the \emph{degree distribution},
for each of these $73$ automorphism orbits, we count the number of nodes
touching a particular graphlet at a node belonging to a particular orbit.
For example, we count how many nodes touch one triangle (i.e., graphlet $G_2$),
how many nodes touch two triangles, how many nodes touch three triangles etc.
We need to separate nodes touching a $G_1$ at an end-node from those touching it at 
a mid-node; thus we count how many nodes touch one $G_1$ at an end-node (i.e., at orbit $1$),
how many nodes touch two $G_1$s at an end-node, 
how many nodes touch three $G_1$s at an end-node etc.
and also how many nodes touch one $G_1$ at a mid-node (i.e., at orbit $2$), how
many nodes touch two $G_1$s at a mid-node, how many nodes touch three $G_1$s at a mid-node etc.
In this way, we obtain $73$ distributions analogous to the degree distribution
(actually, the degree distribution is the distribution for the $0^{th}$ orbit, i.e., for 
graphlet $G_0$).  
Thus, the degree distribution, which has been considered to be a global
network property, is one in the \emph{spectrum} of $73$ 
\emph{``graphlet degree distributions (GDDs)''} measuring local 
structural properties of a network.  Note that GDD is measuring \emph{local} structure,
since it is based on small local network neighborhoods.
The distributions are unlikely to be statistically independent of each other,
although we have not yet worked out the details of the inter-dependence.

\subsection{Network ``GDD Agreement''}\label{sect:gdd_agree}

There are many ways to ``reduce'' the large quantity of numbers representing
$73$ sample distributions.  In this section, we describe one way; there may be 
better ways, and certainly finding better ways to reduce this data is an obvious
future direction.
Some of the details may seem obscure at first; we justify them at the end of this section.

\begin{figure*}[hbtp]
\begin{center}
\begin{tabular}{cc}
{\textsf{(A)} \resizebox{0.45\textwidth}{!}{\includegraphics[angle=270]{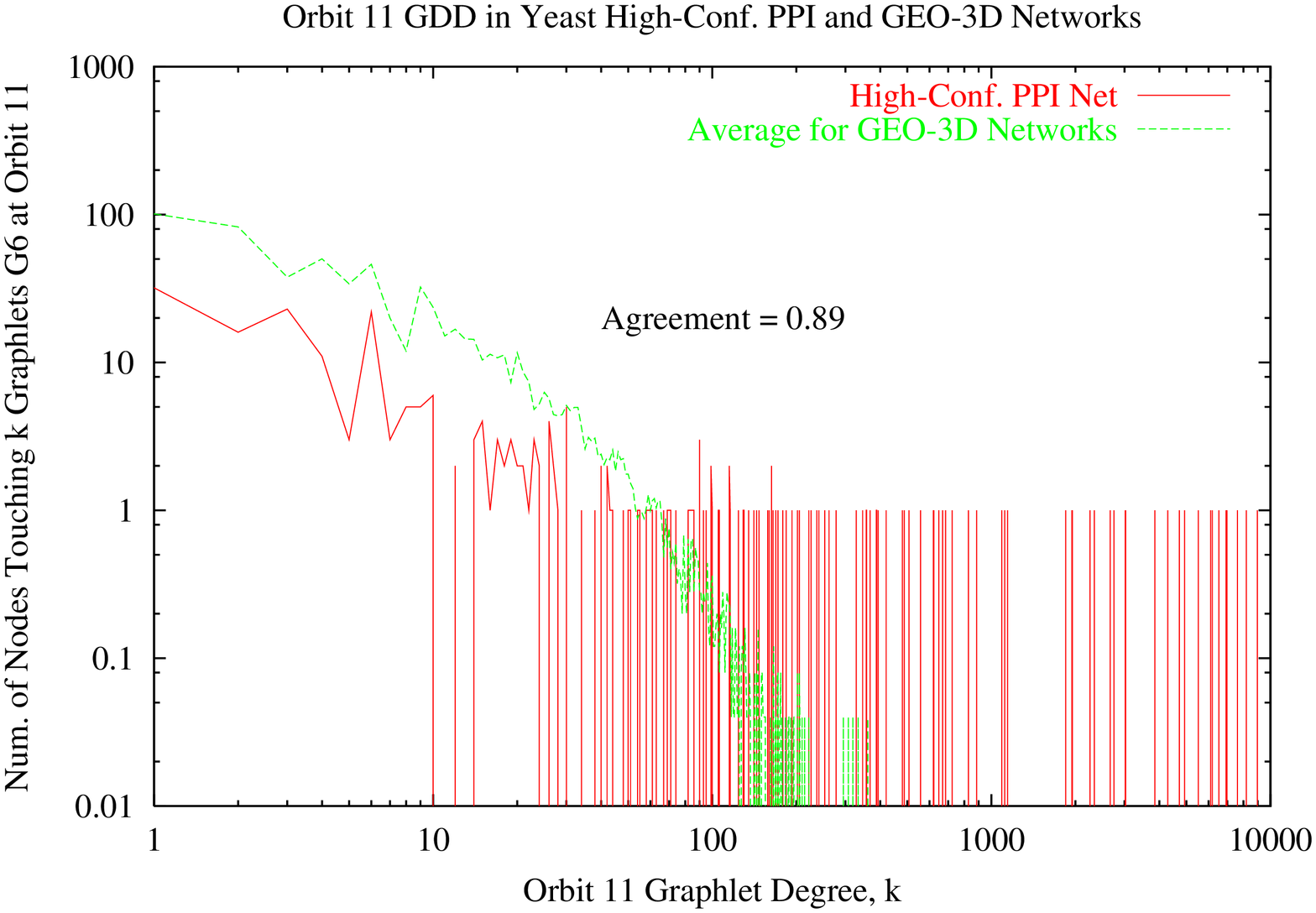}}} &
\textsf{(B)} \resizebox{0.45\textwidth}{!}{\includegraphics[angle=270]{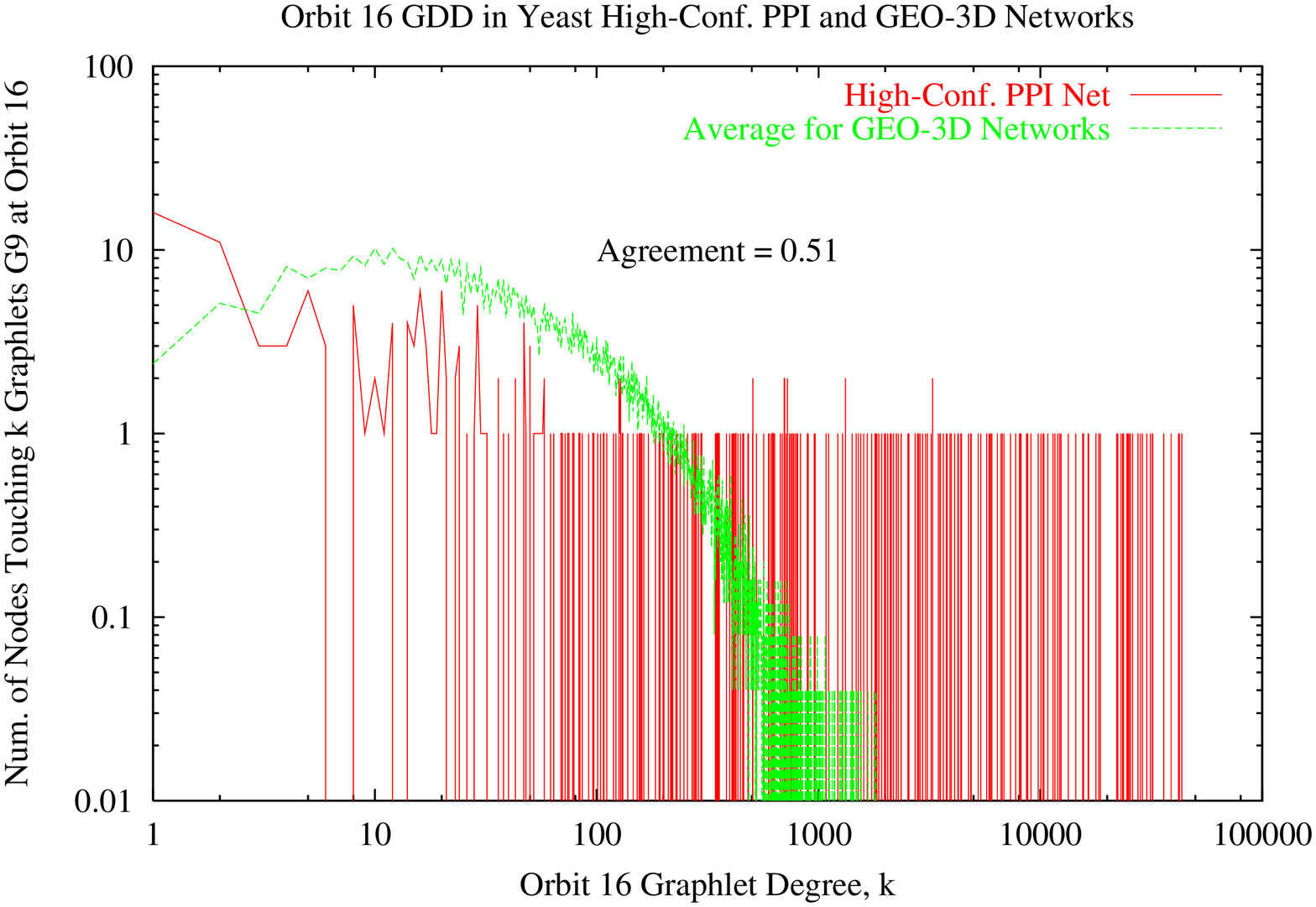}} \\
\end{tabular}
\caption{Examples of graphlet degree distributions (GDDs) for yeast high-confidence PPI network \citep{Mering02} (solid red line) and the average of 25 corresponding 3-dimensional geometric random networks (GEO-3D, dashed green line):
{\bf A. } Orbit 11.
{\bf B. } Orbit 16.
Most counts beyond about $k=20$ are zero, with a few instances of 1 or (very occasionally)
2. This results in a large amount of red and green ink which is mostly noise, as the
distribution fluctuates frequently from 1 to 0 (which is $-\infty$ on our log scale).
The noise could be reduced by applying a broad-band filter, but we have chosen
to leave the data in its raw state, despite the deleterious effect on the
aesthetics of the plot.
}
\label{fig:cl_8_9_b2k_geo_3d}
\end{center}
\end{figure*}

We start by measuring the $73$ graphlet degree distributions (GDDs) 
for each network that we wish to compare.
Let $G$ be a network (i.e., a graph).
For a particular automorphism orbit $j$ (refer to Figure \ref{fig:graphlet_orbits}),
let $d_G^j(k)$ be the sample distribution of the number of nodes in $G$ touching the
appropriate graphlet (for automorphism orbit $j$) $k$ times.  That is, $d_G^j$ represents
the $j^{th}$ graphlet degree distribution (GDD).  
We scale $d_G^j(k)$ as 
\begin{equation} 
S_G^j(k) = \frac{d_G^j(k)}{k}\label{eq:01}
\end{equation}
to decrease the contribution of larger degrees in a GDD (for reasons we describe
later that are illustrated in Figure \ref{fig:cl_8_9_b2k_geo_3d}), and then
normalize the distribution with respect to its total area\footnote{in practice the upper limit of the sum is finite due to finite sample size},
\begin{equation} 
T_G^j = \sum_{k=1}^\infty S_G^j(k).\label{eq:02}
\end{equation}
giving the ``normalized distribution''
\begin{equation} 
N_G^j(k) = \frac{S_G^j(k)}{T_G^j}.\label{eq:03}
\end{equation}
In words, $N_G^j(k)$ is the fraction of the total area under the curve,
over the entire GDD, devoted to degree $k$.  Finally, for two networks
$G$ and $H$ and a particular orbit $j$, we define the ``distance''
$D^j(G,H)$ between their normalized $j^{th}$ distributions as
\begin{equation} 
D^j(G,H) = \left(\sum_{k=1}^\infty[N_G^j(k)-N_H^j(k)]^2\right)^{\frac{1}{2}},\label{eq:04}
\end{equation} 
where again in practice the upper limit of the sum is finite due to the finite
sample.
The distance is between $0$ and $1$, where $0$ means that $G$ and $H$
have identical $j^{th}$ GDDs, and $1$ means that their $j^{th}$ GDDs
are far away.  Next, we reverse $D^j(G,H)$ to obtain the $j^{th}$
\emph{GDD agreement}:
\begin{equation} 
A^j(G,H) = 1 - D^j(G,H),\label{eq:05}
\end{equation} 
for $j \in \{0,1,\ldots,72\}$.  Finally, the {\bf \emph{agreement}}
between two networks $G$ and $H$ is either the arithmetic (equation \ref{eq:06})
or geometric (equation \ref{eq:07}) mean of $A^j(G,H)$ over all $j$, i.e.,
\begin{equation}
A_{arith}(G,H) = \frac{1}{73} \sum_{j=0}^{72}A^j(G,H),\label{eq:06}
\end{equation} 
and
\begin{equation}
A_{geo}(G,H) = \left(\prod_{j=0}^{72}A^j(G,H)\right)^{\frac{1}{73}}.\label{eq:07}
\end{equation}

Now we give the rationale for designing the agreement measure in this way.
There are many different ways to design a measure of agreement between two 
distributions.  They are all heuristic, and thus one needs to examine the
data to design the agreement measure that works best for a particular application.
The justification of our choice of the graphlet degree distribution agreement measure
can be illustrated by an example of two GDDs for the yeast high-confidence PPI 
network \citep{Mering02} and the corresponding 3-dimensional geometric random
networks presented in Figure \ref{fig:cl_8_9_b2k_geo_3d}.  This Figure gives an 
illustration of the GDDs of orbit 11 of the 
PPI and average GDD of orbit 11 in 25 model networks (panel A) being ``closer'' 
than the GDDs of orbit 16 (panel B);
this is accurately reflected by our agreement measure which gives an agreement
of $0.89$ for orbit 11 GDDs and of $0.51$ for orbit 16.  
However, note that the sample distributions extend in the $x$ axis out to degrees
of $10^4$ or even $10^5$; we believe that most of the ``information'' in the
distribution is contained in the lower degrees and that the information in the
extreme high degrees is noise due to bio-technical false positives caused by
auto-activators or sticky proteins \citep{Vidal05}.  
However, without scaling by $1/k$ as in equation
(\ref{eq:01}), both the area under the curve (\ref{eq:02}) and the distance
(\ref{eq:04}) would be dominated by the counts for large $k$.  This explains
the scaling in equation (\ref{eq:01}).  The ``normalization'', equation
(\ref{eq:03}), in performed in order to force both distributions to have a
total area under the curve of 1 before they are compared.  We can now
compute, for each value of $k$, the ``distance'' between two distributions
at that value of $k$.  Formally $k$ is unbounded but in practice it is
finite due to the finite size of the graph.  We then treat the vector of
distances as a vector in the unit cube of dimension equal to the maximum
value of $k$.  We compute the Euclidean distance between two
of these vectors, representing two networks, in equation (\ref{eq:04}).
Finally, we choose to switch from ``distance'' to ``agreement'' in equation
(\ref{eq:05}) simply because we feel agreement is a more intuitive measure.

To gauge the quality of this agreement measure, we computed the average agreements 
between various model (i.e., theoretical) networks.  For example, when
comparing networks of the same type (ER vs ER, ER-DD vs ER-DD,
GEO-3D vs GEO-3D, or SF vs SF),  we found the mean agreement to be 0.84 with
a standard deviation of 0.07.  To verify that
our ``agreement'' measure can give low values for networks that are very different,
we also constructed a ``straw-man'' model graph
called a \emph{circulant}, and compared it to some actual PPI network data.
A circulant graph is constructed by adding ``chords'' 
to a \emph{cycle} on $n$ nodes (examples of cycles on $3$, $4$, and $5$ nodes are 
graphlets $G_2$, $G_5$, and $G_{15}$, respectively) so that $i^{th}$ node on the 
cycle is connected to the $[(i + j) \mod n]^{th}$ and $[(i - j) \mod n]^{th}$ node on 
the cycle.  Clearly, a large circulant with an equal number of nodes edge density as 
the data would not be very representative of a PPI network, and indeed we find that
the agreement between such a circulant, with chords defined by $j \in \{6,12\}$,
and the data is under 0.08.  Note that in most of the fourteen PPI networks, the number of 
edges is abut $3$ times the number of nodes, so we chose circulants with three times 
as many edges as nodes; also, we chose $j>5$ to maximize the number of 3, 4, and 5-node 
graphlets that do not occur in the circulant, since all of the 3, 4, and 5-node
graphlets occur in the data.

\section{Results and Discussion}\label{sect:results}

\begin{figure*}[hbtp]
\begin{center}
\begin{tabular}{cc}

{\textsf{(A)} \resizebox{0.45\textwidth}{!}{\includegraphics[angle=270]{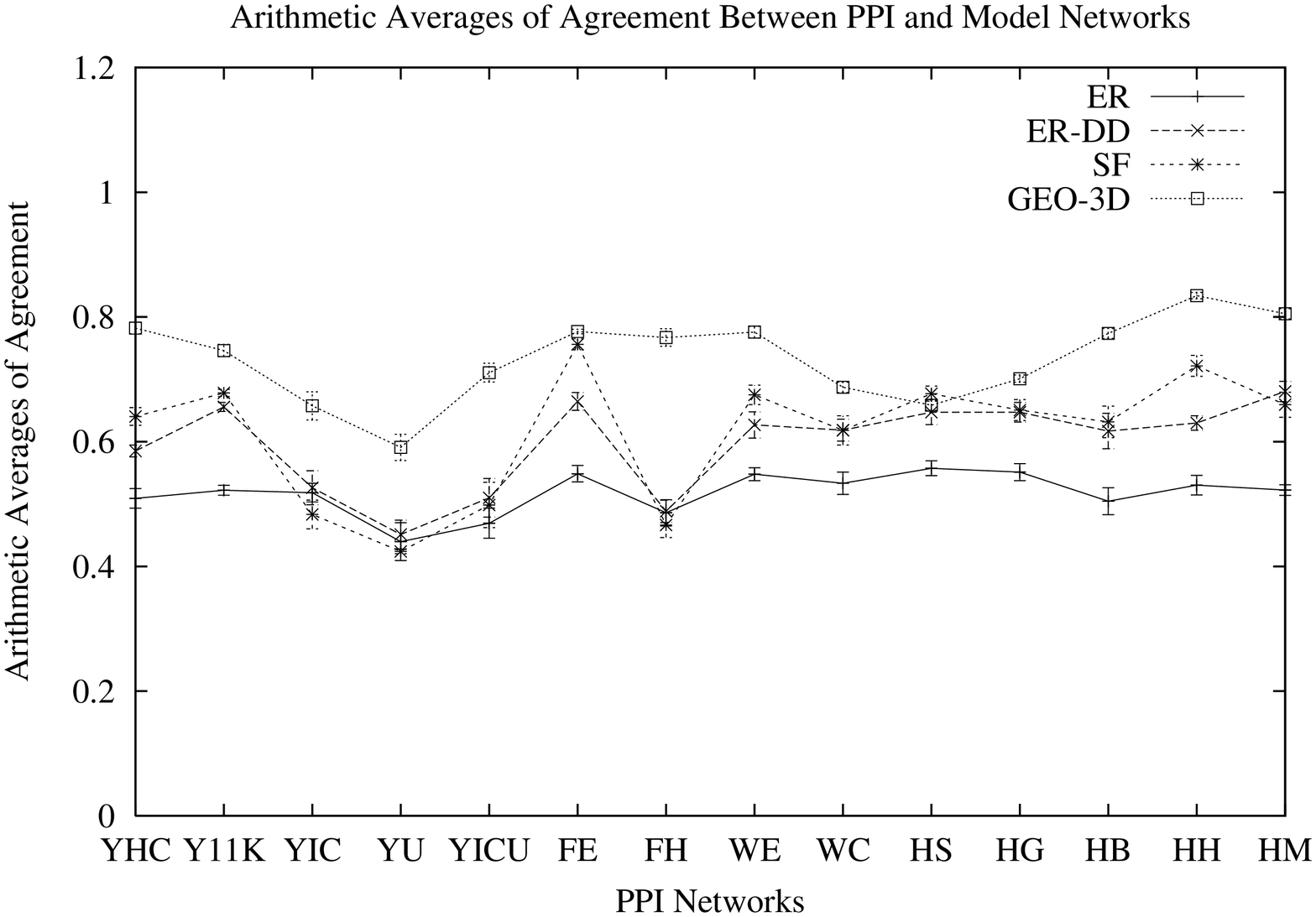}}} &
\textsf{(B)} \resizebox{0.45\textwidth}{!}{\includegraphics[angle=270]{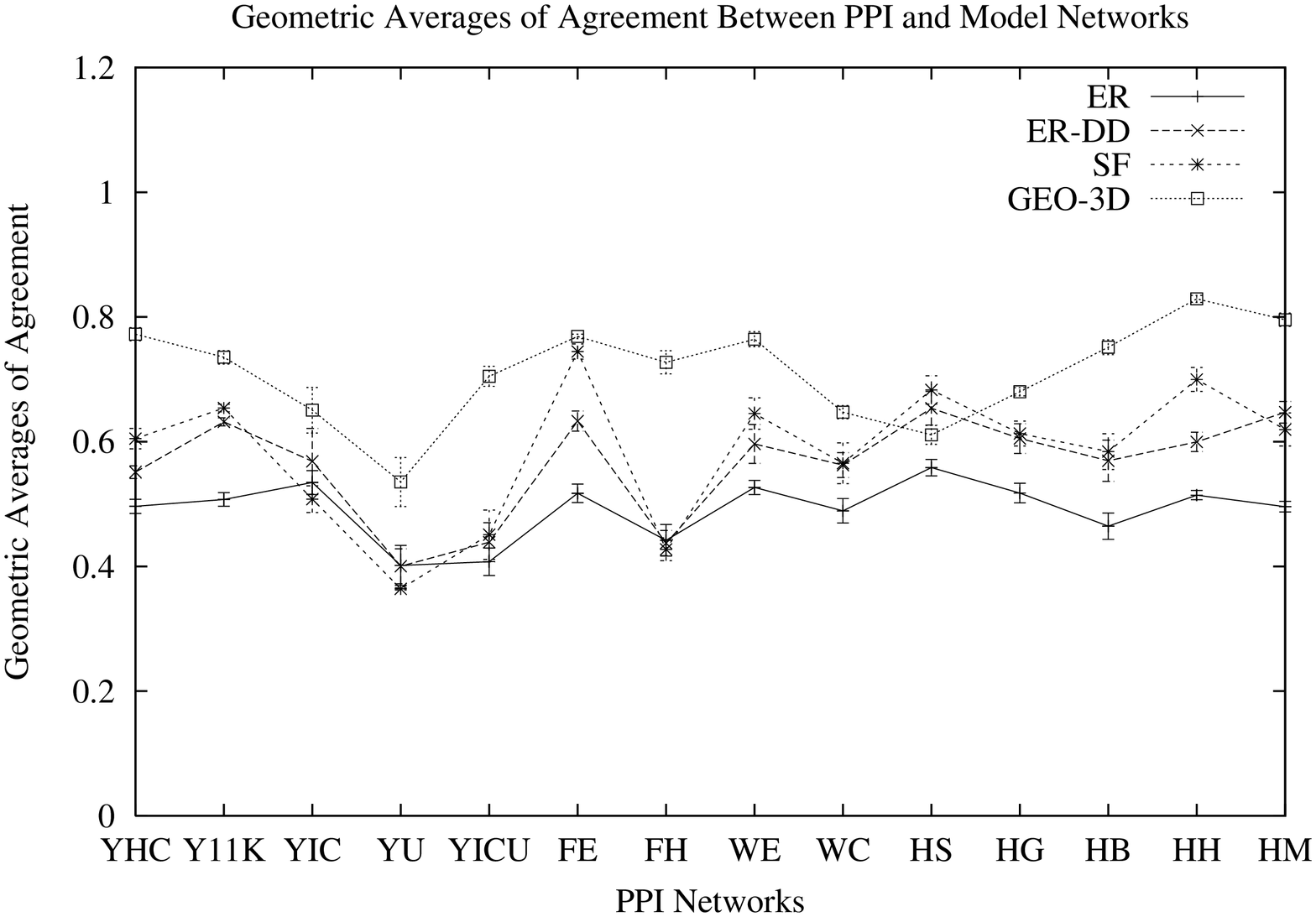}} \\
\end{tabular}
\caption{Agreements between the fourteen PPI networks and their corresponding model networks.  Labels on the horizontal axes are described in section \ref{sect:data_model}.
Averages of agreements between $25$ model networks and the corresponding PPI network are presented for each random graph model and each PPI network, i.e., at each point in the Figure; the error bar around a point is one standard deviation below and above the point (in some cases, error bars are barely visible, since they are of the size of the point).  As described in section \ref{sect:gdd_agree}, the agreement between a PPI and a model network is based on the:
{\bf A. } arithmetic average of $j^{th}$ GDD agreements;
{\bf B. } geometric average of $j^{th}$ GDD agreements.
}
\label{fig:agree_a_avg_11_ppi_5_models}
\end{center}
\end{figure*}

We present the results of applying the newly introduced ``agreement'' 
measure (section \ref{sect:gdd_agree}) to fourteen eukaryotic PPI networks and their 
corresponding model networks of four different network model types (described in 
section \ref{sect:data_model}).  The results show that 3-dimensional
geometric random graphs have exceptionally high agreement with all of the fourteen
PPI networks.

We undertook a large-scale scientific computing task by implementing
the above described new methods and using them to compare agreements across
the four random graph models of fourteen real PPI networks.
Using these new methods, we analyzed a total of $1,414$ networks: fourteen eukaryotic PPI
networks of varying confidence levels described in section \ref{sect:data_model}
and $25$ model networks per random graph model corresponding to each of the 
fourteen PPI networks, where random graph models were ER, ER-DD, SF, and GEO-3D
(described in section \ref{sect:data_model}).  
The largest of these networks had around $7,000$ nodes and over $20,000$ edges.
For each of the fourteen PPI networks and each of the four random graph models, 
we computed averages and standard deviations 
of graphlet degree distribution (GDD) \emph{agreements} 
between the PPI and the $25$ corresponding model networks belonging to 
the same random graph model.  The results are presented in Figure 
\ref{fig:agree_a_avg_11_ppi_5_models}.

Erdos-Renyi random graphs (ER) show about $0.5$ agreement with each
of the PPI networks while scale-free networks of type ER-DD and SF show a slightly 
improved agreement (ER-DD networks are random scale-free, since the degree 
distributions forced on them by the corresponding PPI networks roughly follow power law).  
Note that GEO-3D networks show the highest agreement 
for \emph{all} but one of the fourteen PPI networks 
(Figure \ref{fig:agree_a_avg_11_ppi_5_models}).  For HS PPI network, it is
not clear which of the GEO-3D, SF, and ER-DD models agrees the most with the data, 
since the average agreements of HS network with these models are about the same and within one
standard deviation from each other.  GEO-3D and SF model are similarly tied for the FE network.
Since networks belonging to the same random graph model have average agreement of 
$0.84$ with a standard deviation of $0.07$ (shown in section \ref{sect:gdd_agree}), 
the agreements of over $0.7$, that most of the PPI networks have
with the GEO-3D model, are very good.  Note that eight out of the fourteen
PPI networks have agreements with GEO-3D model of over $0.75$; since networks of
the same type agree on average by $0.84 \pm 0.07$, we conclude that the agreements 
of $0.75$ are exceptionally high and are unlikely to be beaten by another network 
model under this measure.  Also, it is interesting that GEO-3D model shows high agreement
with PPI networks obtained from various experimental techniques (Y2H, TAP, HMS-PCI)
as well as from human curation (see section \ref{sect:data_model}).
Note that this does not mean that GEO-3D is the best possible model.  For example,
it may be possible to construct a different ``agreement'' measure that is more
sensitive and under which a model better than GEO-3D may be apparent.  However,
we believe that the current ``agreement'' measure is sensitive and meaningful 
enough to conclude that GEO-3D is a better model than ER, ER-DD, and SF.

\section{Conclusion}

We have constructed a new measure of structural similarity between large networks 
based on the graphlet degree distribution.  The degree 
distribution is the first one in the sequence of graphlet degree distributions 
that are constructed in a structured and systematic way to impose a large number
of constraints on the structure of networks being compared.  This new measure is
easily extendible to a greater number of constraints simply by adding more
graphlets to those in Figure \ref{fig:graphlet_orbits}, although this would
add significantly to the cost of computing agreements; the extensions 
are limited only by the available CPU.
Based on this new network similarity measure, we have shown that almost all of the fourteen eukaryotic
PPI networks resulting from various high-throughput experimental techniques, as well as
curated databases, are better modeled by geometric random graphs than by Erdos-Renyi,
random scale-free, or Barabasi-Albert scale-free networks.

\section*{Acknowledgement}

We thank Derek Corneil and Wayne Hayes for helpful comments and discussions,
Pierre Baldi for providing computing resources, and Jason Lai for help with programming.

\end{document}